\newif\ifsubmode
\submodefalse

\newif\ifprintfig
\printfigtrue

\newif\ifemulate
\emulatetrue

\ifsubmode
\documentclass[12pt,preprint]{aastex}
\received{}
\accepted{}
\journalid{}
\articleid{}
\else
   \documentclass{emulateapj}
   \submitted{{\it Accepted for publication in ApJ}}
\fi
\usepackage{multirow}
\usepackage{amsmath,amssymb,mathrsfs}

\newcommand{\like}{\mathscr{L}}
\newcommand{\bftheta}{\mathbf{\Theta}}

\def\spose#1{\hbox to 0pt{#1\hss}}
\def\simlt{\mathrel{\spose{\lower 3pt\hbox{$\mathchar"218$}}
     \raise 2.0pt\hbox{$\mathchar"13C$}}}
\def\simgt{\mathrel{\spose{\lower 3pt\hbox{$\mathchar"218$}}
     \raise 2.0pt\hbox{$\mathchar"13E$}}}

\shorttitle{Are Ultra-Faints Cusps?}
\shortauthors{Zolotov~et~al.}

\begin{document} 

\title{Are the ultra-faint dwarf galaxies just cusps?}

\author{Adi\ Zolotov\altaffilmark{1},David\ W.\ Hogg\altaffilmark{1},Beth\ Willman\altaffilmark{2},}

\altaffiltext{1}{Center for Cosmology and Particle Physics, Department of Physics, New York University, 4 Washington Place, New York, NY 10003; az481@nyu.edu}
\altaffiltext{2}{Haverford College, Department of Astronomy, 370 Lancaster Avenue, Haverford, PA 19041}

\date{Dec 9, 2010}
\begin{abstract}

  We develop a technique to investigate the possibility that some of
    the recently discovered ultra-faint dwarf satellites of the Milky
    Way might be cusp caustics rather than gravitationally self-bound
    systems. Such cusps can form when a stream of stars folds,
    creating a region where the projected 2-D surface density is
    enhanced. In this work, we construct a Poisson maximum likelihood test to
    compare the cusp and exponential models of any substructure on an
    equal footing. We apply the test to the Hercules dwarf (d
    $\sim$ 113 kpc, M$_V \sim -6.2$, $e \sim 0.67$).  The flattened
    exponential model is strongly favored over the cusp model in the
    case of Hercules, ruling out at high confidence that Hercules is a
    cusp catastrophe. This test can be applied to any of the Milky Way
    dwarfs, and more generally to the entire stellar halo population,
    to search for the cusp catastrophes that might be expected in an
    accreted stellar halo.
\end{abstract}

\keywords{Galaxy --- halo; galaxies --- dwarfs}
          
\section{INTRODUCTION}\label{intro_sec}
In the last five years, more than a dozen ultra-faint Milky Way
satellites have been discovered with one millionth of the Milky Way's
luminosity or less. These objects have stars with [Fe/H] at least as
low as -3.0, [Fe/H] spreads of up to 1 dex
\citep{kirby08a,frebel10a,simon10a}, and velocity dispersions of $\sim
3 - 7$ km sec$^{-1}$ \citep{Simon2007, Martin2007}, and scale sizes of
30 - 200 pc \citep{Martin2008}. A common explanation for these
observed properties thus far is that we are observing dwarf galaxy
member stars moving on bound orbits within highly dark matter
dominated potential wells.  Within this scenario, the velocity
dispersions of these objects have been interpreted to possibly imply
mass-to-light ratios of $10^{2 - 4}$ within their central 300 pc
\citep{strigari08a} and that galaxies can have total luminosities of
only $\sim 1000 L_{\odot}$ - far less than the luminosities of some
individual stars.

These remarkable conclusions should require a high burden of proof:
Are there any other plausible explanations for the ultra-faint Milky
Way satellites as a population?  One possibility is that the
ultra-faints could have intrinsic near-zero velocity dispersions (and
thus no dark matter), but that binary stars have inflated their
observed dispersions. While this scenario could be conceivable for a
small number of objects, it cannot explain away the entire ultra-faint
galaxy population \citep{simon10a, Martinez2010,
mcconnachie10a}. Moreover, the low amount of and large spread in their
iron abundances show that these objects cannot have formed in the same
way as globular clusters, ruling out (disrupting) globular cluster
scenarios.  The possible inflated ellipticities of ultra-faint dwarfs
relative to their more luminous neighbors \citep{Martin2008} combined
with claims of irregular isophotes tempts a conclusion that these are
tidal remnants.  However, claims of isophote irregularity have been
inconclusive for the majority of ultra-faint satellites. Recently,
\citet{Martin2010} have found that the shape of the Hercules dwarf
might be explained by tidal disruption.

A fourth possibility is that (a subset of) the kinematically cold,
spatial overdensities that have been identified as ultra-faint
galaxies are instead cusp caustics of cold stellar debris.  Cold
substructures are both observed \citep[][e.g.]{Grillmair2006, Yanny2003,
Belokurov2006} and predicted to be abundant in the Milky Way's
halo. Shell-like structures can form during the disruption of a low
mass satellite, as the satellite interacts with a massive galaxy, and
have been seen in both simulations \citep{Hernquist1988} and observations
\citep{Malin1983} of early-type galaxies. After an interaction creates such
a sheet of stars, the sheet can evolve and fold in ways that produce
visible fold caustics (``catastrophes''), cusp caustics, and
higher-order caustics in the observed two-dimensional distribution of
stars \citep{Tremaine1999}. A cusp caustic in a two-dimensional
observable space appears as a highly elliptical centroid of stars
where two fold caustics meet, creating an asymmetrical flare-like
feature. In this scenario, some of the ultra-faint dwarfs may actually
be cusps and some of the cold streams could be folds.

In this paper, we develop a technique for testing the hypothesis that
ultra-faint dwarf galaxies are the 2-D projections of such folded stellar
sheets.  We apply this technique to the Hercules Milky Way satellite,
which has d $\sim$ 113 kpc, $M_V \sim$ -6.2, and $e \sim$ 0.67
\citep{Sand2009, Coleman2007}. We choose this object because it has the highest
ellipticity of any of the Milky Way's dwarfs and is thus a good
candidate for the cusp model.

\section{Observational Data}
\citet[][hereafter C07]{Coleman2007} obtained Large Binocular
Telescope imaging of the Hercules dwarf galaxy in the Gunn $r$ band
(25 minutes), $V$ band (20 minutes), and $B$ band (30 minutes) to
derive its structural parameters. The Large Binocular Camera spans a
23$'$ x 23$'$ field of view. C07 applied color-color-magnitude
filtering in colors $c_1$ and $c_2$ (combinations of $r$, $V$,
and $B$) to their point source catalog to select objects most
consistent with belonging to Hercules rather than to the field.  In
this paper we utilize the $\sim$10,000 stars from the C07 CMD selected
subset of their LBT/LBC Hercules catalog.  This catalog is identical
to what they used to derive the structure of Hercules.  Please see
C07 for further details on this data set.

\section{Model Comparison: Cusp vs Exponential}
We aim to determine whether a cusp model is a better explanation of
the spatial distribution of the elongated Hercules dwarf than a
flattened exponential model. To perform this model comparison, we
specify each model explicitly, compute the Poisson likelihood of the
data given the models, and use the maximum likelihood of each model to
determine which best fits the data.

\subsection{Constructing the Models}
Both of the exponential and cusp models tested here have six free
parameters. The cusp model's free parameters are: central RA and Dec,
rotation angle and scale, a smoothing scale, and a constant background
level. The exponential model's free parameters are: central RA and
Dec, a rotation angle, ellipticity, half-light radius, and a constant
background level. For each model we step through the six parameters
($\mathbf{\Theta}$) and evaluate the likelihood of the data given each
model. The full likelihood, which is a product of the individual star
likelihoods, is:

\begin{align}
\like(\mathbf{\Theta})= \prod_{i=1}^{N_{star}} \frac{1}{Z(\mathbf{\Theta})}\Sigma(x_i|\bftheta)
\end{align}
\begin{align}
Z(\mathbf{\Theta})=\frac{1}{N_{random}} \sum_{j=1}^{N_{random}} \Sigma(x_{random,j}|\mathbf{\Theta}) 
\end{align}
where $\mathbf{\Theta}$ is the list of parameters for the cusp or
exponential model, $\Sigma(\cdot)$ is the surface density of the model
given the parameters, and $Z(\mathbf{\Theta})$ is a normalization
constant, estimated using a spatially randomized stellar catalog
over the observed sky region. $N_{star}$ is the total number of stars,
and $\mathbf{x}_i$ is the two-dimensional position (on the sky) of
star $i$. $N_{random}$ is the total number of random points, set to
equal $N_{star}$, and $\mathbf{x}_{random,j}$ is the position of
random point $j$.

\subsubsection{The Cusp  Model}

The cusp surface density model is computed following the work of
\citet{Tremaine1999}. First, a background-free $\Sigma_{cusp}$ is
computed on a grid in the natural coordinate system $\mathbf{\xi}$:
\begin{align}
\Sigma_{cusp}(\xi_1,\xi_2)=\sum_{roots} [1+\frac{1+y^2}{(\xi_2 +3y^2)^2}]^{1/2},
\end{align}
where 
\begin{align}
\xi_1=-y^3-\xi_2y .
\end{align}
In three dimensions, coordinates $\xi_1$ and $\xi_2$ make up the 2-D
observable space, while coordinate y is the third, and hidden,
dimension. As a third degree polynomial, equation (4) has either one
or three solutions, depending on the position in
$\mathbf{\xi}$. Equation (3) is to be summed over all the possible
roots of variable y, obtained by solving equation (4). In regions
where there is no cusp, there will only be one solution to equation
(4), and hence no summation in the calculation of the surface density
is necessary. On the other hand, in regions of $\mathbf{\xi}$ where
there is a cusp, the variable y will be triple valued, and equation
(3) must be summed over all three roots. Because this function has a
large dynamic range, we compute it on a fine grid of $1000 \times
1000$ pixels. Even with such fine pixels, however, the value of the
cusp function can vary significantly within a pixel in regions of the
image where the cusp is present. At such pixels, we evaluate the
function (equation 3) at multiple random locations within the pixel.
This adaptive sampling sets the number of samples per pixel to a
constant multiple (10) of the value of the function. This essentially
ensures that we are sampling more in the most interesting regions of
the image, where the cusp is present. $\Sigma_{cusp}$ has no free
parameters in the natural coordinate system.

After the cusp image has been computed on a grid, it is smoothed with
a symmetric two-dimensional Gaussian kernel with variance
$\sigma^2$. We have computed seven cusp models, each with a different
smoothing scale, with values of $\sigma=[1,2,4,8,16,32,64]$
pixels. The two dimensional positions $\mathbf{x_i}$ (RA,Dec) of each
star $i$ on the sky is converted to a position $\mathbf{\xi_i}$ in the
natural coordinate system of the cusp model by a shift, scale and
rotation:%
\begin{align}
 \mathbf{\xi_i}=\mathbf{R}\cdot [\mathbf{x_i}-\mathbf{X}],
\end{align} %
where $\mathbf{R}$ is a $2 \times 2$ matrix encoding the two free parameters of
scale and rotation, and $\mathbf{X}$ represents the two free
parameters that shift the position of the center of the model on the
sky. Finally the surface density $\Sigma$ computed at point 
$\mathbf{\xi_i}$ is the sum of the background free 
$\Sigma_{cusp}$, and a constant background level $\Sigma_{bg}$:
\begin{align}
\Sigma(\mathbf{\xi})=\Sigma_{cusp}(\mathbf{\xi})+\Sigma_{bg} .
\end{align}

\subsubsection{The Exponential Model}

We follow the same procedure for the exponential model as for the cusp
model. The exponential  model, in its natural units, is:
\begin{align}
\Sigma_{exp}(\mathbf{\xi})=exp(-|\mathbf{\xi}|),
\end{align}
and the model is not smoothed at all. The $2 \times 2$ matrix $\mathbf{R}$ used
in this case is a symmetric tensor of scale and shear with three free
parameters, used to calculate the position angle ($\theta$), ellipticity
($\epsilon$), and scale radius ($r_h$) of the model.

\bigbreak

All models have been defined on grids of $1000 \times 1000$ pixels, on
the interval $\mathbf{\xi}$=[-5,5], and have six degrees of
freedom. Figure 1 shows the surface brightness grids for cusp models
with $\sigma=[1,2,4,8]$ pixels and for the exponential model, each in their
natural units.

\begin{table}[t]
\begin{tabular}{lll|l}
\multicolumn{3}{c}{Maximum Likelihood Parameters} \\
\hline
\hline
Parameter & Measured & Uncertainty & $\ln \like$\\ \hline
\multicolumn{3}{c}{Best Exponential Profile} \\\hline
R.A. (h m s)&16:31:03.9 &$8''$ \\
DEC (d m s)&+12:47:10 &$7''$\\
$r_h$ (arcmin)& 5.3 &0.4 & 382\\
$\epsilon$ & 0.65 &0.03\\
$\theta$ (degrees) & -75.6 &2.1 \\\hline
\multicolumn{3}{c}{Best Cusp Profile} \\ \hline
R.A. (h m s)&16:31:30 & $ 17''$\\
DEC (d m s)&+12:46:15 &$ 4''$\\
scale\tablenotemark{a}  & 37.6 &2.1& 280\\%
\tablenotetext{a}{arcmin per coordinate distance measured in $\mathbf{\xi}$}%
$\theta$ (degrees) & -78.1 & $ 3.1$ \\
$\sigma$ (pixels) & 2\tablenotemark{b} & -- \\\hline
\hline
\tablenotetext{b}{equivalent to $\sigma=0.75$ arcmin}
\end{tabular}
\end{table}

\subsection{MCMC}
We use a Markov Chain Monte Carlo (MCMC) method to calculate the
maximum likelihood of each model, and its corresponding
parameters. MCMC simulates the likelihood for a set of parameters by
sampling from the posterior distribution through a series of random
draws. We specifically use the Metropolis-Hastings Algorithm, which
generates a random walk and steps through each parameter in such a way
that more probable values of parameter space are more often stepped
to. For both the cusp and exponential models we choose a separable
prior that is flat in center position ($\mathbf{X}$), flat in the
components of the $\mathbf{R}$ matrix, and flat in the background
level ($\Sigma_{bg}$), with a lower limit of zero. There were no
derivatives in our prior at the peaks in the likelihood, so the choice
of prior does not affect any of our results. The acceptance ratios for
all parameters are $\sim 0.5$. Though less computationally efficient,
this method is similar to the maximum likelihood technique used by
\citep{Martin2008} to determine the structural parameters of dwarf
galaxies, and is better at escaping local minima.

\subsection{Results}
We find that a flattened exponential model is a far better fit to the
Hercules data than any of the seven cusp models tested in this
work. Of the seven cusp models tested, the cusp with a smoothing scale
of $\sigma=2$ pixels ($0.75$ arcmin) was found to be the most
probable. The difference between the maximum likelihood of the
exponential and best fit cusp model is 
\begin{align}
\like_{expo}/ \like_{cusp} = e^{-102} .
\end{align}
Nominally this rules out the best cusp model in favor of the
exponential at a confidence level of $>99.99\%$.

Table 1 lists the parameters at maximum
likelihood for the exponential and best cusp models.  The structural
parameters for the exponential model yield a position angle for
Hercules at $-76^{\circ}$, and an ellipticity of 0.65 (defined as
$\epsilon=1-b/a$), both in good agreement with the findings of
\citet{Coleman2007}, \citet{Martin2008}, and \citet{Sand2009}.
The parameters that maximize the likelihood of the cusp model are such
that the model is slightly smoothed, and highly scaled.  This is not
surprising since the cusp model is a poor fit to the data, and at its
optimal configuration it will be smoothed and scaled in such a way to
resemble an flattened exponential-like distribution near the
center. If Hercules were a 2-D projection of a cusp, the spatial
distribution of its stars would show a highly elongated center of
stars, along with two ``tails'' indicative of the two fold
catastrophes. Figure 2 shows the predicted spatial distribution of
stars for the best fit exponential and cusp models.

While the cusp model is strongly ruled out in favor of a flattened
exponential, we make no claims that the exponential profile
is the best available model for Hercules. 

\begin{figure*}
\epsscale{1.0}
\plotone{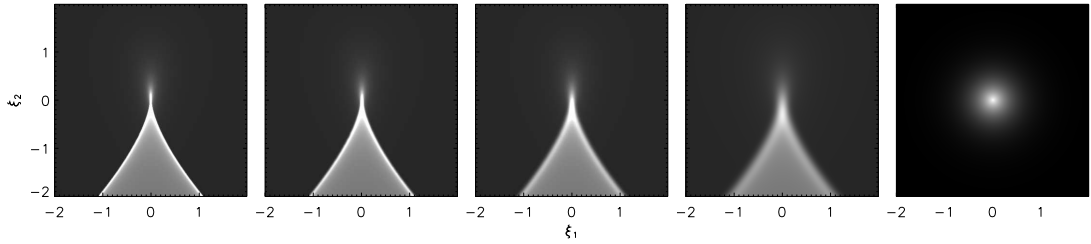}
\caption{
From left to right: The surface brightness maps produced by cusp
models smoothed with a Gaussian kernel of variance $\sigma=[1,2,4,8]$
pixels and by an exponential model, as described in Section 3.1. Each
was calculated on a 1000 x 1000 pixel grid, and is plotted in its
natural coordinate system, $\mathbf{\xi} $. }
\end{figure*}

\section{Conclusion \& Discussion}
It has been suggested that some of the ultra-faint dwarf satellites of
the Milky Way may be the 2-D projection of cusp caustics, and not
gravitationally bound galaxies. The highly elliptical Hercules dwarf
is a good candidate for testing the cusp model, which predicts a
highly elongated centroid of stars. We have established a hypothesis
test in order to determine whether the Hercules overdensity is better
modeled with a cusp or an exponential, and rule out at very high
confidence the possibility that Hercules is a cusp catastrophe. While
other surface brightness models have been fit to Hercules data, such
as the King \citep{King1966} and Plummer \citep{Plummer1911} profiles,
we have chosen to not test such models here, and so have not evaluated
how they compare to the flattened exponential model. This work also
has not evaluated if the properties of the Hercules galaxy are better
explained by the exponential equilibrium model than by models of tidal
disruption \citep[e.g][]{Martin2010}. Our intention in this work is
not to advocate that an exponential profile is the best explanation of
the observed ellipticity of Hercules, but rather to definitively
remove from consideration the possibility that Hercules is a cusp
caustic.

The work outlined in this paper can be used to test any of the Milky
Way dwarfs suspected of being cusps. The highly elliptical dwarfs Ursa
Major I and II are interesting candidates for such a test, though both
lack obvious evidence of the double spatial tail indicative of the
fold caustics that meet at the cusp. Furthermore, the cusp caustic
predicts triple-valued velocities for stellar members of the
cusp. Spectroscopic data of the ultra-faint dwarfs suspected of being
cusps can thus be used to further rule out or confirm such
hypotheses. For example, spectroscopic data in the vicinity of the
ultra-faint galaxy Segue I \citep{Niederste2009, simon10a} show a lot
of interesting structure, and could be used to test if Segue I is a
cusp.

The techniques introduced in this paper can be further extended to
implement a search for cusp and fold caustics on the entire sky, using
data from a large survey like the Sloan Digital Sky Survey (SDSS),
\textsl{Gaia}, or the Large Synoptic Survey Telescope (LSST). 
Stellar streams and dwarf galaxies, which can be thought of,
respectively, as one and (blurry) zero dimensional objects in phase
phase, have been found with multi-band two-dimensional angular maps in
photometric surveys, such as SDSS. Such data sets are two-dimensional,
though some distance information can be determined at low
signal-to-noise using colors, and are well suited for finding low
dimensional structures. Given the number of streams already known in
the Milky Way's stellar halo, it is likely that higher dimensional
structures in six-dimensional phase space exist, but have not yet been
found in current data \citep{Tremaine1999}. Future surveys that return
full (or nearly full) phase-space information, such as the
\textsl{Gaia} mission
\citep{Perryman2001}, will be much better suited than current surveys
for finding systems with higher dimensionality. Meanwhile, such structures may
only be easily identifiable in photometric data if their projection
into 2-D observable space creates regions of enhanced density, like
the cusp and fold caustics.

\begin{figure*}
\epsscale{1.0}
\plotone{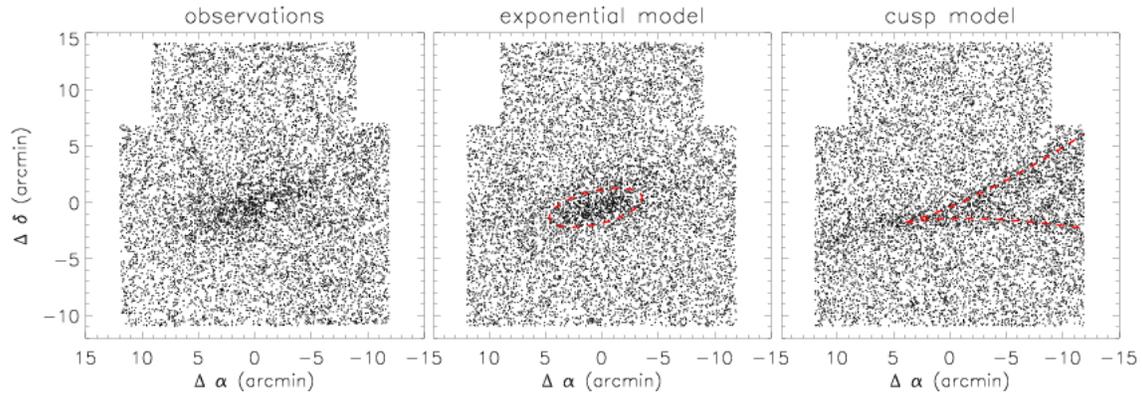}
\caption{
Left Panel- The observed data set of Hercules stars from
\citet{Coleman2007}. The elongated spatial distribution of this
overdensity has raised questions as to whether it is a bound dwarf
galaxy or a cusp fold. Middle Panel- The predicted spatial
distribution of stars in the exponential model. Right Panel- The
predicted spatial distribution of stars in the cusp model with the
highest likelihood (smoothed with $\sigma=2$ pixels). The dashed red
lines mark the best fit exponential and cusp models in middle and left
panels, respectively.}
\end{figure*}

\acknowledgements

We thank Niayeshi Afshordi for inspiring this paper, and thank Matt
Coleman for sharing the catalog of data used in
\citet{Coleman2007}. We thank the anonymous referee for helpful comments. 
We also thank James Bullock for helpful comments on an earlier draft
of this work. AZ and BW thank NSF AST-090844 for support.  BW also
thanks NSF AST-0908193 for support. DWH was partially supported by
NASA NNX08AJ48G and the NSF AST-0908357.


\begin{thebibliography}{22}
\expandafter\ifx\csname natexlab\endcsname\relax\def\natexlab#1{#1}\fi

\bibitem[{{Belokurov} {et~al.}(2006){Belokurov}, {Zucker}, {Evans}, {Gilmore},
  {Vidrih}, {Bramich}, {Newberg}, {Wyse}, {Irwin}, {Fellhauer}, {Hewett},
  {Walton}, {Wilkinson}, {Cole}, {Yanny}, {Rockosi}, {Beers}, {Bell},
  {Brinkmann}, {Ivezi{\'c}}, \& {Lupton}}]{Belokurov2006}
{Belokurov}, V., {Zucker}, D.~B., {Evans}, N.~W., {Gilmore}, G., {Vidrih}, S.,
  {Bramich}, D.~M., {Newberg}, H.~J., {Wyse}, R.~F.~G., {Irwin}, M.~J.,
  {Fellhauer}, M., {Hewett}, P.~C., {Walton}, N.~A., {Wilkinson}, M.~I.,
  {Cole}, N., {Yanny}, B., {Rockosi}, C.~M., {Beers}, T.~C., {Bell}, E.~F.,
  {Brinkmann}, J., {Ivezi{\'c}}, {\v Z}., \& {Lupton}, R. 2006, \apjl, 642,
  L137

\bibitem[{{Coleman} {et~al.}(2007){Coleman}, {de Jong}, {Martin}, {Rix},
  {Sand}, {Bell}, {Pogge}, {Thompson}, {Hippelein}, {Giallongo}, {Ragazzoni},
  {DiPaola}, {Farinato}, {Smareglia}, {Testa}, {Bechtold}, {Hill}, {Garnavich},
  \& {Green}}]{Coleman2007}
{Coleman}, M.~G., {de Jong}, J.~T.~A., {Martin}, N.~F., {Rix}, H., {Sand},
  D.~J., {Bell}, E.~F., {Pogge}, R.~W., {Thompson}, D.~J., {Hippelein}, H.,
  {Giallongo}, E., {Ragazzoni}, R., {DiPaola}, A., {Farinato}, J., {Smareglia},
  R., {Testa}, V., {Bechtold}, J., {Hill}, J.~M., {Garnavich}, P.~M., \&
  {Green}, R.~F. 2007, \apjl, 668, L43

\bibitem[{{Frebel} {et~al.}(2010){Frebel}, {Simon}, {Geha}, \&
  {Willman}}]{frebel10a}
{Frebel}, A., {Simon}, J.~D., {Geha}, M., \& {Willman}, B. 2010, \apj, 708, 560

\bibitem[{{Grillmair}(2006)}]{Grillmair2006}
{Grillmair}, C.~J. 2006, \apjl, 645, L37

\bibitem[{{Hernquist} \& {Quinn}(1988)}]{Hernquist1988}
{Hernquist}, L., \& {Quinn}, P.~J. 1988, \apj, 331, 682

\bibitem[{{King}(1966)}]{King1966}
{King}, I.~R. 1966, \aj, 71, 64

\bibitem[{{Kirby} {et~al.}(2008){Kirby}, {Simon}, {Geha}, {Guhathakurta}, \&
  {Frebel}}]{kirby08a}
{Kirby}, E.~N., {Simon}, J.~D., {Geha}, M., {Guhathakurta}, P., \& {Frebel}, A.
  2008, \apjl, 685, L43

\bibitem[{{Malin} \& {Carter}(1983)}]{Malin1983}
{Malin}, D.~F., \& {Carter}, D. 1983, \apj, 274, 534

\bibitem[{{Martin} {et~al.}(2008){Martin}, {de Jong}, \& {Rix}}]{Martin2008}
{Martin}, N.~F., {de Jong}, J.~T.~A., \& {Rix}, H. 2008, \apj, 684, 1075

\bibitem[{{Martin} {et~al.}(2007){Martin}, {Ibata}, {Chapman}, {Irwin}, \&
  {Lewis}}]{Martin2007}
{Martin}, N.~F., {Ibata}, R.~A., {Chapman}, S.~C., {Irwin}, M., \& {Lewis},
  G.~F. 2007, \mnras, 380, 281

\bibitem[{{Martin} \& {Jin}(2010)}]{Martin2010}
{Martin}, N.~F., \& {Jin}, S. 2010, ArXiv e-prints

\bibitem[{{Martinez} {et~al.}(2010){Martinez}, {Minor}, {Bullock},
  {Kaplinghat}, {Simon}, \& {Geha}}]{Martinez2010}
{Martinez}, G.~D., {Minor}, Q.~E., {Bullock}, J., {Kaplinghat}, M., {Simon},
  J.~D., \& {Geha}, M. 2010, ArXiv e-prints

\bibitem[{{McConnachie} \& {C{\^o}t{\'e}}(2010)}]{mcconnachie10a}
{McConnachie}, A.~W., \& {C{\^o}t{\'e}}, P. 2010, \apjl, 722, L209

\bibitem[{{Niederste-Ostholt} {et~al.}(2009){Niederste-Ostholt}, {Belokurov},
  {Evans}, {Gilmore}, {Wyse}, \& {Norris}}]{Niederste2009}
{Niederste-Ostholt}, M., {Belokurov}, V., {Evans}, N.~W., {Gilmore}, G.,
  {Wyse}, R.~F.~G., \& {Norris}, J.~E. 2009, \mnras, 398, 1771

\bibitem[{{Perryman} {et~al.}(2001){Perryman}, {de Boer}, {Gilmore}, {H{\o}g},
  {Lattanzi}, {Lindegren}, {Luri}, {Mignard}, {Pace}, \& {de
  Zeeuw}}]{Perryman2001}
{Perryman}, M.~A.~C., {de Boer}, K.~S., {Gilmore}, G., {H{\o}g}, E.,
  {Lattanzi}, M.~G., {Lindegren}, L., {Luri}, X., {Mignard}, F., {Pace}, O., \&
  {de Zeeuw}, P.~T. 2001, \aap, 369, 339

\bibitem[{{Plummer}(1911)}]{Plummer1911}
{Plummer}, H.~C. 1911, \mnras, 71, 460

\bibitem[{{Sand} {et~al.}(2009){Sand}, {Olszewski}, {Willman}, {Zaritsky},
  {Seth}, {Harris}, {Piatek}, \& {Saha}}]{Sand2009}
{Sand}, D.~J., {Olszewski}, E.~W., {Willman}, B., {Zaritsky}, D., {Seth}, A.,
  {Harris}, J., {Piatek}, S., \& {Saha}, A. 2009, \apj, 704, 898

\bibitem[{{Simon} \& {Geha}(2007)}]{Simon2007}
{Simon}, J.~D., \& {Geha}, M. 2007, \apj, 670, 313

\bibitem[{{Simon} {et~al.}(2010){Simon}, {Geha}, {Minor}, {Martinez}, {Kirby},
  {Bullock}, {Kaplinghat}, {Strigari}, {Willman}, {Choi}, {Tollerud}, \&
  {Wolf}}]{simon10a}
{Simon}, J.~D., {Geha}, M., {Minor}, Q.~E., {Martinez}, G.~D., {Kirby}, E.~N.,
  {Bullock}, J.~S., {Kaplinghat}, M., {Strigari}, L.~E., {Willman}, B., {Choi},
  P.~I., {Tollerud}, E.~J., \& {Wolf}, J. 2010, ArXiv e-prints

\bibitem[{{Strigari} {et~al.}(2008){Strigari}, {Bullock}, {Kaplinghat},
  {Simon}, {Geha}, {Willman}, \& {Walker}}]{strigari08a}
{Strigari}, L.~E., {Bullock}, J.~S., {Kaplinghat}, M., {Simon}, J.~D., {Geha},
  M., {Willman}, B., \& {Walker}, M.~G. 2008, \nat, 454, 1096

\bibitem[{{Tremaine}(1999)}]{Tremaine1999}
{Tremaine}, S. 1999, \mnras, 307, 877

\bibitem[{{Yanny} {et~al.}(2003){Yanny}, {Newberg}, {Grebel}, {Kent},
  {Odenkirchen}, {Rockosi}, {Schlegel}, {Subbarao}, {Brinkmann}, {Fukugita},
  {Ivezic}, {Lamb}, {Schneider}, \& {York}}]{Yanny2003}
{Yanny}, B., {Newberg}, H.~J., {Grebel}, E.~K., {Kent}, S., {Odenkirchen}, M.,
  {Rockosi}, C.~M., {Schlegel}, D., {Subbarao}, M., {Brinkmann}, J.,
  {Fukugita}, M., {Ivezic}, {\v Z}., {Lamb}, D.~Q., {Schneider}, D.~P., \&
  {York}, D.~G. 2003, \apj, 588, 824

\end{thebibliography}

\end{document}